\documentclass[preprint,5p,numbers,sort&compress]{elsarticle}

\usepackage{graphicx}
\usepackage{dcolumn}
\usepackage{bm}
\usepackage{longtable}
\usepackage{float}
\usepackage{threeparttable}
\usepackage[colorlinks = true, citecolor=blue,urlcolor=blue,linkcolor=blue]{hyperref}
\usepackage{color} 
\usepackage{booktabs}

\usepackage{supertabular}%
\usepackage{float}

\usepackage{url}
\usepackage{array}
\usepackage{amssymb}
\usepackage{dblfloatfix}
\usepackage{tabu}

\begin{document}


\title{Single neutron transfer on $^{23}$Ne and its relevance for the pathway of nucleosynthesis in astrophysical X-ray bursts} 

\author[sur]{G. Lotay\corref{cor1}}
\ead{g.lotay@surrey.ac.uk}
\cortext[cor1]{Corresponding author}

\author[sur,tri]{J. Henderson}

\author[sur]{W. N. Catford}

\author[gue,sul]{F. A. Ali}

\author[tri]{J. Berean}

\author[tri,ubc]{N. Bernier\fnref{fn1}}
\fntext[fn1]{Present address: Department of Physics and Astronomy, University of the Western Cape, P/B X17, Bellville, ZA-7535, South Africa}

\author[tri]{S. S. Bhattacharjee\fnref{fn2}}
\fntext[fn2]{Present address: Institute of Experimental and Applied Physics, Czech Technical University in Prague, Husova 240/5, 110 00 Prague 1, Czech Republic}

\author[tri]{M. Bowry\fnref{fn3}}
\fntext[fn3]{Present address: School of Computing, Engineering and Physical Sciences, University of the West of Scotland, Paisley PA1 2BE, United Kingdom}

\author[tri]{R. Caballero-Folch}

\author[tri,sfu]{B. Davids}

\author[tor]{T. E. Drake}

\author[tri]{A. B. Garnsworthy}

\author[gue]{F. Ghazi Moradi}

\author[tri]{S. A. Gillespie\fnref{fn4}}
\fntext[fn4]{Present address: National Superconducting Cyclotron Laboratory, Michigan State University, East Lansing, Michigan 48824, USA}

\author[gue]{B. Greaves}

\author[tri]{G. Hackman}

\author[sur]{S. Hallam}

\author[gue]{D. Hymers}

\author[gue]{E. Kasanda}

\author[tri]{D. Levy}

\author[ten]{B. K. Luna}

\author[tri]{A. Mathews}

\author[ohi]{Z. Meisel}

\author[sur]{M. Moukaddam\fnref{fn5}}
\fntext[fn5]
{Present Address: Universit{'e} de Strasbourg, IPHC, 23 rue du Loess, 67037 Strasbourg, France}

\author[tri,gue]{D. Muecher}

\author[tri]{B. Olaizola\fnref{fn6}}
\fntext[fn6]{Present Address: ISOLDE, CERN, CH-1211 Geneva 23, Switzerland}

\author[lpc]{N. A. Orr}

\author[tri]{H. P. Patel}

\author[ten]{M. M. Rajabali}

\author[tri,ubc]{Y. Saito}

\author[tri]{J. Smallcombe\fnref{fn7}}
\fntext[fn7]
{Present Address: University of Liverpool, Liverpool, L69 3BX, UK.}

\author[sur,tri]{M. Spencer}

\author[gue]{C.\ E.\ Svensson}

\author[sfu]{K. Whitmore}

\author[tri,yor]{M. Williams}

\address[sur]{Department of Physics, University of Surrey, Guildford, GU2 7XH, United Kingdom }
\address[tri]{TRIUMF, Vancouver, British Columbia, V6T 2A3, Canada}

\address[gue]{Department of Physics, University of Guelph, Guelph, Ontario N1G 2W1, Canada}
\address[sul]{Department of Physics, College of Education, University of Sulaimani, P.O. Box 334, Sulaimani, Kurdistan Region, Iraq}

\address[ubc]{Department of Physics and Astronomy, University of British Columbia, Vancouver, BC V6T 1Z4, Canada}

\address[sfu]{Department of Physics, Simon Fraser University, Burnaby, British Columbia V5A 1S6, Canada}

\address[tor]{Department of Physics, University of Toronto, Toronto, Ontario M5S 1A7, Canada}

\address[ten]{Department of Physics, Tennessee Technological University, Cookeville, Tennessee 38505, USA}

\address[ohi]{Department of Physics $\&$ Astronomy, Ohio University, Athens, Ohio 45701, USA}

\address[lpc]{LPC-Caen, IN2P3/CNRS, UniCaen, ENSICAEN, Normandie Universit{\'e}, 14000 Caen, France}

\address[yor]{Department of Physics, University of York, Heslington, York, YO10 5DD, United Kingdom}




\date{\today}

\begin {abstract}

We present new experimental measurements of resonance strengths in the astrophysical $^{23}$Al($p,\gamma$)$^{24}$Si reaction, constraining the pathway of nucleosynthesis beyond $^{22}$Mg in X-ray burster scenarios. Specifically, we have performed the first measurement of the ($d,p$) reaction using a radioactive beam of $^{23}$Ne to explore levels in $^{24}$Ne, the mirror analog of $^{24}$Si. Four strong single-particle states were observed and corresponding neutron spectroscopic factors were extracted with a precision of $\sim$20$\%$. Using these spectroscopic factors, together with mirror state identifications, we have reduced uncertainties in the strength of the key $\ell$ = 0 resonance at $E_r$ = 157 keV, in the astrophysical $^{23}$Al($p,\gamma$) reaction, by a factor of 4. Our results show that the $^{22}$Mg($p,\gamma$)$^{23}$Al($p,\gamma$) pathway dominates over the competing $^{22}$Mg($\alpha,p$) reaction in all but the most energetic X-ray burster events ($T>0.85$ GK), significantly affecting energy production and the preservation of hydrogen fuel.

\end {abstract}


\maketitle

Type-I X-ray bursts represent thermonuclear explosions on the surfaces of accreting neutron stars in close binary systems \cite{Wallace,Schatz,Schatz2}. They exhibit dramatic, recurrent increases in luminosity and constitute the most frequent stellar eruptions to occur in our Galaxy. In between bursts, energy is generated at a constant rate by the $\beta$-limited hot CNO cycles \cite{Caughlan,Audouze}. However, as the temperature of the accreted material increases, the triple-$\alpha$ reaction becomes favourable, igniting the burst, and nucleosynthesis proceeds along the proton-rich side of stability via the $\alpha$$p$ process \cite{Fisker} [a series of ($p,\gamma$) and ($\alpha,p$) reactions], and the $rp$ process \cite{Schatz2} [a series of ($p,\gamma$) reactions and $\beta$$^{+}$ decays], ending in the Sn-Te mass region.

Recently, advances in computing power have allowed for detailed models of X-ray burst nucleosynthesis to be constructed \cite{Woosley,Paxton,Jose,Fisker}, incorporating complex reaction networks and hundreds of nuclear species ranging from stable isotopes up to the proton drip line. Strikingly, despite the vast number of reactions included, only a handful of nuclear processes have been highlighted as having a noticeable effect on the observational properties of X-ray bursts \cite{Parikh,Cyburt,Meisel}. In particular, the $^{23}$Al($p,\gamma$)$^{24}$Si reaction, which permits flow beyond masses of $A=22$ in the early phases of the $rp$ process, is postulated to have a strong influence on the inferred surface gravitational redshift ($1+z$) \cite{Meisel}. The redshift is directly related to the neutron star compactness \cite{Meisel2} and thus, any experimental constraints placed on the $^{23}$Al($p,\gamma$) reaction rate will help to reveal new facets of the underlying compact objects involved. Furthermore, at the $^{22}$Mg, $rp$-process waiting point, the $^{22}$Mg($p,\gamma$)$^{23}$Al($p,\gamma$) reaction sequence is expected to compete significantly with the $^{22}$Mg($\alpha,p$) reaction \cite{Randhawa}, affecting the overall energy generation in X-ray bursters. Specifically, a prevailing $^{22}$Mg($p,\gamma$)$^{23}$Al($p,\gamma$) pathway results in less energetic burning during the burst rise, preserving hydrogen for later burning and extending the burst tail. The exceptional measurements now available for the structure of burst light curves \cite{Galloway,Galloway2} are amenable to confront simulations of the burst explosions.





Previous studies of the $^{23}$Al($p,\gamma$) reaction \cite{Schatz3,Yoneda,Banu,Wolf} indicate that the rate is dominated by resonant capture on the 5/2$^{+}$ ground state of $^{23}$Al to excited states above the proton-emission threshold energy of 3292(19) keV in $^{24}$Si \cite{Wang}. However, the strengths of these resonances remain uncertain, due to the scarcity of experimental data. Most recently, Wolf {\it et al.} utilised the $^{23}$Al($d,n$) reaction to investigate the properties of excited states in $^{24}$Si \cite{Wolf}. In that study \cite{Wolf}, $\gamma$ decays were observed from three excited states, including the key $\ell$ = 0, proton-unbound resonant level at 3449(5) keV, which is expected to have the most significant influence on the $^{23}$Al($p,\gamma$) reaction over the temperature range of X-ray bursts. Moreover, by measuring angle-integrated cross sections of excited levels in $^{24}$Si, Wolf {\it et al.} were able to place the first constraints on proton spectroscopic factors, reducing uncertainties in both the direct and resonant capture components of the $^{23}$Al($p,\gamma$) reaction \cite{Wolf}. However, the absolute values of spectroscopic factors reported in Ref. \cite{Wolf} carry large uncertainties of order 60$\%$ because their extraction relied on the use of shell-model calculations to determine the relative contributions of multiple $\ell$-transfers. For example, where states are populated by a mixture of $\ell$ = 0 and $\ell$ = 2 transfer, the work of Ref. \cite{Wolf} was forced to use the ratio of strengths predicted by the shell model, but it has been pointed out \cite{Lotay} that the shell model consistently fails to predict this ratio correctly. Consequently, the rate of the $^{23}$Al($p,\gamma$) reaction is still weakly constrained over the temperature range of Type-I X-ray bursts and a more robust experimental measurement is demanded.

%
%

A direct measurement of the $^{23}$Al($p,\gamma$) reaction is not presently feasible. As such, any further experimental constraints must rely on indirect techniques. In this regard, several studies have shown that precise evaluations of proton capture reactions may be achieved via the concept of isospin \cite{Iliadis,Pain,Margerin}. Specifically, neutron spectroscopic factors of excited states in mirror nuclei, that correspond to analogs of ($p,\gamma$) resonances, can be used to accurately determine the strengths of resonances governing the rate of stellar reactions in explosive astrophysical environments \cite{Iliadis,Pain,Margerin}. In this Letter, we present a first experimental measurement of the $^{23}$Ne($d,p$) transfer reaction to study excited states in $^{24}$Ne. These levels correspond to $T=2$, mirror analogs of key resonant states in the $^{23}$Al($p,\gamma$)$^{24}$Si reaction. By coupling the TIGRESS $\gamma$-array \cite{TIGRESS} to the SHARC charged-particle detection system \cite{SHARC}, neutron spectroscopic factors were extracted to a precision of $\sim$20$\%$. This reduces uncertainties in $^{23}$Al + $p$ resonance strengths by a factor $\sim$4 and, hence, defines the relative importance of the $^{22}$Mg($p,\gamma$)$^{23}$Al($p,\gamma$) and $^{22}$Mg($\alpha,p$) reaction sequences over the temperature range of X-ray bursts.


\begin{table*}[h]
\centering
\begin{threeparttable}

\caption{Properties of excited states in the $T=2$, $A=24$ system, as determined in the present work and reported in earlier literature. Excitation energies are given in keV and shell-model spectroscopic factors were determined using the USDA interaction \cite{Brown}. In Ref. \cite{Hoffman}, no uncertainties for $E_x$ ($^{24}$Ne) are given, but we expect $\leq$ 2 keV based on HPGe calibration. The present $C^2S_{(d,p)}$ is the summed $\ell$ = 2 strength, extracted assuming transfer to $0d_{5/2}$ (errors and limits, see text). For comparison, $C^2S_{SM}$ is the sum of USDA shell-model values for $0d_{5/2}$ and $0d_{3/2}$.}

\renewcommand{\tabcolsep}{1.2pc}

\begin{tabular}{cccccccc}

\hline
 $E_{x}$ ($^{24}$Ne) \cite{Hoffman} \tnote{a} &  $J^{\pi}$ & $\ell_p$ & $C^2S_{(d,p)}$ & $C^2S_{SM}$ & $C^2S_{(d,n)}$ \cite{Wolf} & Analog State in $^{24}$Si \cite{Wolf} \\ 
 
\\ 
\hline
 
0 & 0$^{+}$ & 2 & 3.42(68)  & 3.50 & $\leq2.8$ & 0 \\
1981 &  2$^{+}$ & 0 & 0.28(6) & 0.28  & 0.6(2) & 1874(3)\\
& & 2 & 0.37(7) & 0.19 & 0.4(1) & \\
3871 & 2$^{+}$ & 0 & 0.44(9)  & 0.42  & 0.7(4) & 3449(5) \\
& & 2 & 0.23(5)  & 0.17 & 0.3(2) & \\
3962 & 4$^{+}$ & 2 & $\leq0.012$ & 0.014  & 0.07(4) & 3471(6)\\
4765 & 0$^{+}$ & 2 & $\leq0.19$  & 0.21 & 0.8(4) &  4170\tnote{b} \\
4886 & 3$^{+}$ & 0 & 0.56(11) & 0.58 & & 4470\tnote{b} \\
&  & 2 & $\leq$ 0.19 &  0.17 & &  \\
\hline

\label{table1}
\end{tabular}
\begin{tablenotes}
\item[a]{Excitation energy uncertainties are not provided in Ref. \cite{Hoffman} but are assumed to be $\sim$1 keV, based on the observation of $\gamma$-ray transitions.}
\item[b]{Taken from theoretical calculations of Ref. \cite{Herndl}}
\end{tablenotes}
\end{threeparttable}
\end{table*}

A beam of radioactive $^{23}$Ne$^{2+}$ ions was accelerated to 8.0 MeV/nucleon and an intensity of $\sim$2 $\times$ 10$^{4}$ pps, by the ISAC-II facility at TRIUMF and bombarded a 1 mg/cm$^2$ (CD$_2$)$_n$ foil for 93 hrs. Prompt $\gamma$ rays were recorded using the TIGRESS array of 12 Compton-suppressed HPGe detectors \cite{TIGRESS}, while charged particles including protons from the $^{23}$Ne($d,p$) reaction were measured in the SHARC silicon array \cite{SHARC}. Beyond the target, 40 cm downstream, the TRIFOIL detector \cite{Wilson,Matta} was placed (a 20 $\mu$m foil of BC400 plastic scintillator viewed by three photomultiplier tubes and mounted behind a passive stopper foil of 110 $\mu$m Al). The TRIFOIL setup (a) stopped the $^{23}$Ne beam and $^{24}$Ne reaction products in the scintillator and counted them, (b) stopped the $^{23}$Na beam contaminant ($\sim$40$\%$ of the beam) and $^{24}$Na reaction products in the Al foil so that they had no TRIFOIL tag and (c) also in the Al, stopped fusion-evaporation products from reactions on carbon in the $CD_2$ target. The TRIFOIL also gave a direct measurement of the average counting rate of the beam over the entire 93 hours of data acquisition to a precision of $<$ 1$\%$. The rejection of $^{23}$Na-induced events was verified by the complete removal of $^{24}$Na $\gamma$-ray peaks when imposing the TRIFOIL requirement. The beam composition was also measured at regular intervals using a Bragg ionization detector \cite{Bragg} and background from other contaminant isobars was found to be negligible. Energy and efficiency calibrations were performed using standard $\gamma$-ray ($^{152}$Eu and $^{60}$Co) and charged-particle (triple alpha) sources. The absolute normalisation was determined using the measured number of incident $^{23}$Ne ions, the target thickness and the H:D ratio, as determined from elastic scattering around $\theta_{cm}$ = 50$^{\circ}$ measured simultaneously throughout the acquisition.

\begin{figure}[!htbp]
  \includegraphics[width=\linewidth]{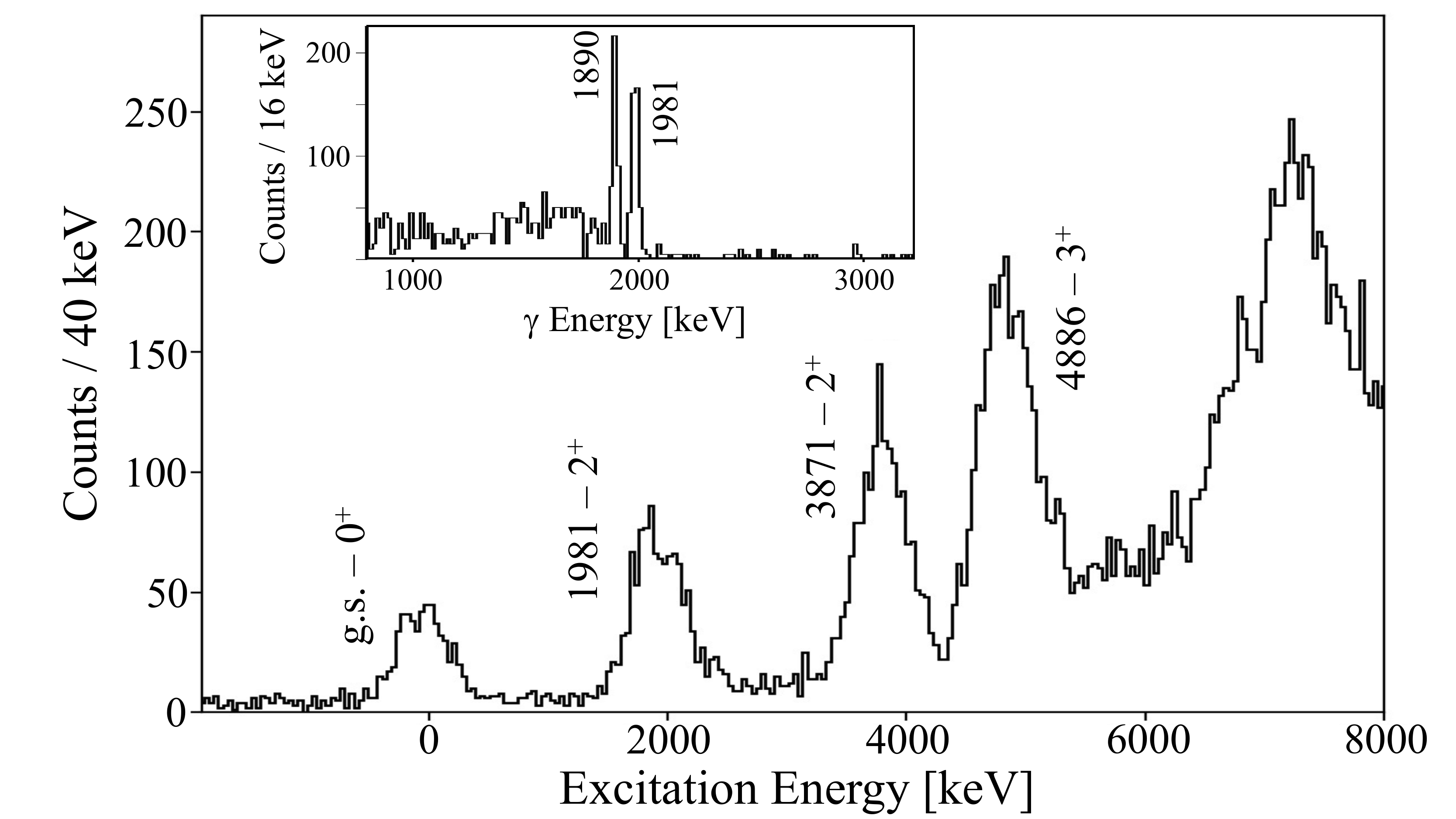}%
\caption{Excitation energy spectrum obtained following $^{23}$Ne($d,p$) transfer. (Inset) Gamma decays observed with a gate placed on the 3871-keV excitation energy peak.}
\label{Fig1}
\end{figure}

\begin{figure}[!htbp]
  \includegraphics[width=\linewidth]{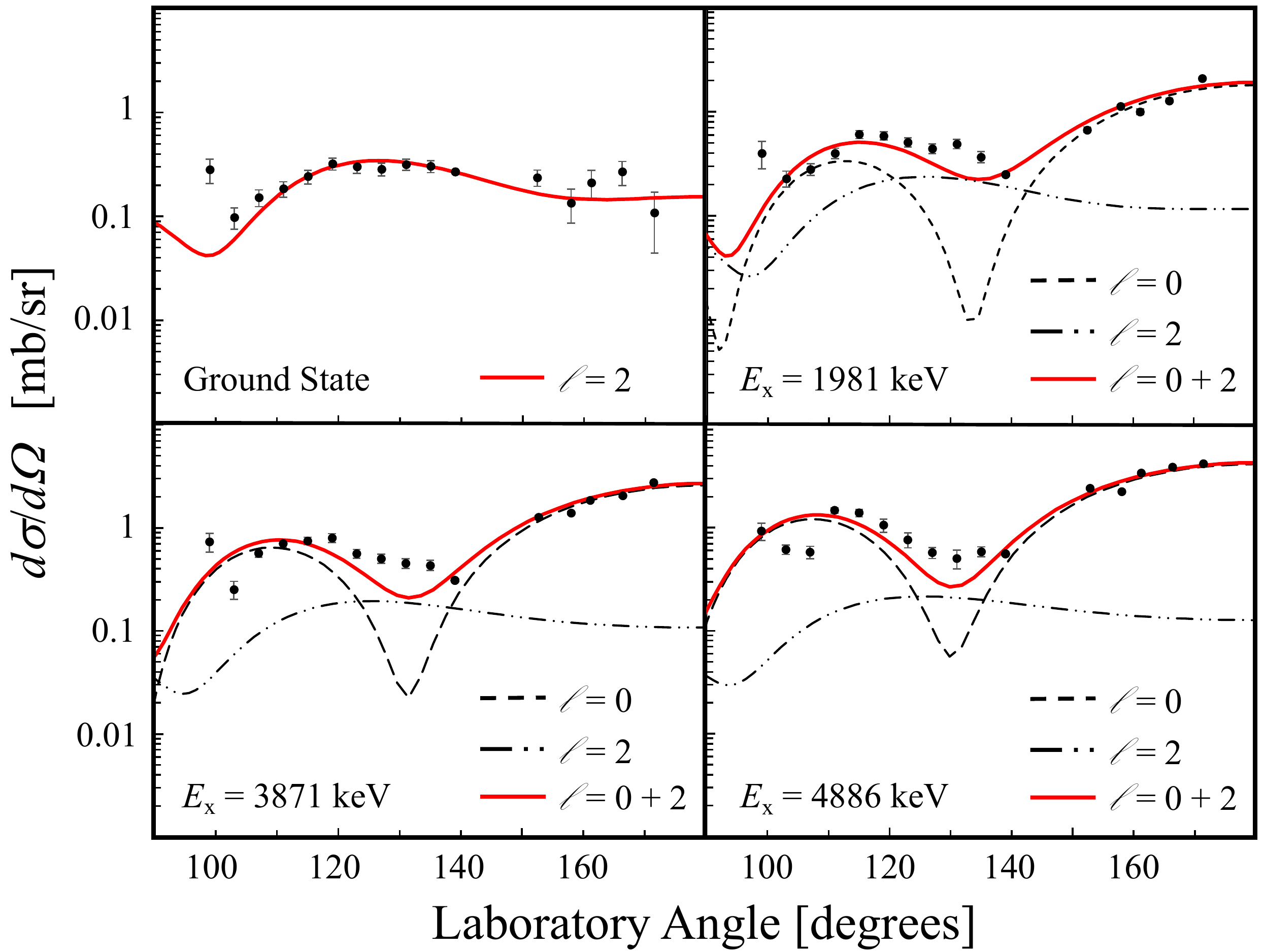}%
\caption{(color online) Angular distributions of protons in the $^{23}$Ne($d,p$) reaction compared with best-fit TWOFNR calculations (statistical error bars only). Data for the 4886-keV state may include the unresolved 4765 keV, 0$^{+}$ state (for the $\ell$ = 2 component only). For details, see text. Uncertainties shown are statistical only.}
\label{Fig2}
\end{figure}

Figure \ref{Fig1} illustrates the excitation energy of states in $^{24}$Ne populated via the ($d,p$) reaction. As can be seen, four strongly populated states are observed at 0, 1981, 3871 and 4886 keV, in good agreement with previously reported 0$^{+}_1$, 2$^{+}_1$, 2$^{+}_2$ and 3$^{+}_1$ levels in $^{24}$Ne \cite{Hoffman} and the theoretical calculations of Ref. \cite{Herndl}. That being said, additional 4$^{+}_1$ and 0$^{+}_2$ excited states are also expected in this energy region in $^{24}$Ne at 3962 and 4765 keV \cite{Hoffman}, respectively, which would not be resolvable using proton detection alone, due to the $\sim$300 keV (FWHM) excitation energy resolution of SHARC. In this regard, the simultaneous detection of $\gamma$ rays is of crucial importance. In particular, by placing gates across the observed energy peaks at 3871 and 4886 keV, and viewing coincident $\gamma$ rays within the TIGRESS array, it was possible to rule out any significant population of the 4$^{+}_1$, 3962-keV and 0$^+_2$, 4765-keV excited states via the $^{23}$Ne($d,p$) reaction. For example, when a gate was placed across the 3871-keV proton peak, we observe 1890- and 1981-keV $\gamma$-ray peaks of equal intensity from the cascade decay from 3871 keV (inset of Fig. 1). The numbers of counts are 76 $\pm$ 10 and 84 $\pm$ 10, respectively. The surplus for the 1981-keV peak is 8 $\pm$ 14 which is consistent with zero and gives a 2$\sigma$ upper limit (allowing for the double counting) of 12$\%$ of the combined population of the two states. This is the basis of the limit on the spectroscopic factor for the 3962-keV state in Table \ref{table1}. Consequently, we conclude that the 3962-keV excited state in $^{24}$Ne was not appreciably populated and, based on an upper limit analysis of 1981-keV transitions originating from the 4$^{+}_1$ state, we set a stringent upper limit on its spectroscopic factor, $C^2S_{(\ell=2)}$ $\leq$ 0.012. This is in agreement with our shell model calculations using NuShellX \cite{Brown2} with the USD-A interaction \cite{Brown}, which predict $C^2S_{(\ell=2)}$ = 0.01 for the 4$^{+}_1$ level in $^{24}$Ne. In contrast, a similar procedure to the above was not possible for the expected 4765/4886-keV doublet due to a considerable level of background in the $\gamma$-ray energy region of interest. Whilst the observed $\ell$ = 0 angular distribution for this doublet, shown in Fig. \ref{Fig2}, may be ascribed entirely to the known 3$^{+}$, 4886-keV level in $^{24}$Ne \cite{Hoffman}, we adopt an upper limit of 0.19 for the $\ell$ = 2 component of the spectroscopic factor for both the 4765- and 4886-keV excited states in $^{24}$Ne.

\begin{table*}[h]
\center
\begin{threeparttable}

\caption{Properties of resonant states in the $^{23}$Al($p,\gamma$) reaction used in the present analysis, together with a comparison to resonance strengths based on spectroscopic factors reported in Ref. \cite{Wolf}. Excitation energies in $^{24}$Si and present $C^2S$ values are as in Table \ref{table1}. These were used to calculate $\Gamma_p$ and (together with the $\Gamma_{\gamma}$ from USDA shell-model calculations, see text) $\omega\gamma$. For states shown in Table \ref{table1} with $\ell$ = 0 and $\ell$ = 2 contributions, only the $\ell$ = 0 is included here since it overwhelmingly dominates the resonance strength. Upper limits have been determined to a 68$\%$ confidence level.}

\renewcommand{\tabcolsep}{0.65pc}
\begin{tabular}{ccccccccc}

\hline
$E_x$, $^{24}$Si & $E_r$ & $J^{\pi}$ & $\ell_p$ & $C^2S$ & $\Gamma_p$ & $\Gamma_{\gamma}$ & Present $\omega\gamma$ & Previous $\omega\gamma$ \tnote{a} \\ 
 (keV) & (keV) &                &               &               & (eV)               & (eV)                              & (eV) & (eV) \\
 
 \hline
 
3449(5) & 157(20) & 2$^{+}$ & 0 & 0.44(9) & 8.2(17) $\times$ 10$^{-5}$ & 1.8 $\times$ 10$^{-2}$ & 3.4(7) $\times$ 10$^{-5}$ & 5.4(31) $\times$ 10$^{-5}$ \\
3471(6) & 179(20) & (4$^{+}$) \tnote{b} & 2 & $\leq0.012$ & $\leq1.0\times10^{-7}$ & 2.5 $\times$ 10$^{-4}$ & $\leq7.6$ $\times$ 10$^{-8}$ & 4.4(26) $\times$ 10$^{-7}$ \\
         &      & (0$^{+}$) \tnote{b} & 2 & $\leq$ 0.19 & $\leq$ 1.6 $\times$ 10$^{-6}$ & 3.0 $\times$ 10$^{-5}$ & $\leq$ 1.3 $\times$ 10$^{-7}$ & 5.6(28) $\times$ 10$^{-7}$ \\
4170 \tnote{c} & 878 \tnote{c} & 0$^{+}$ & 2 & $\leq$ 0.19 & $\leq$ 32 & 1.6 $\times$ 10$^{-4}$  & 1.3 $\times$ 10$^{-5}$ \tnote{d} & - \\
4470 \tnote{c} & 1178 \tnote{c} & 3$^{+}$ & 0 & 0.56(11) &  $3.0\times10^4$ & $8.9\times10^{-3}$   & 5.2 $\times$ 10$^{-3}$ \tnote{d}  & - \\

\hline

\label{table2}
\end{tabular}
\begin{tablenotes}
\item[a]{Previous resonance strengths have been estimated based on spectroscopic factors reported in Ref. \cite{Wolf}}
\item[b]{For the 179-keV resonance, we currently favour a 4$^{+}$ assignment based on mirror energy difference arguments. However, for completeness, we provide resonance strength determinations for both 4$^{+}_1$ and 0$^{+}_2$ assignments.}
\item[c] {Adopted from Ref. \cite{Herndl}}
\item[d] {Resonance strength determination dominated by theoretically calculated $\gamma$-ray partial width}
\end{tablenotes}
\end{threeparttable}
\end{table*}

An angular distribution analysis of the 0-, 1981-, 3871- and 4886-keV excited states in $^{24}$Ne, shown in Fig. \ref{Fig2}, confirms the spin-parity assignments of Ref. \cite{Hoffman}. However, with the exception of the ground state, which necessarily exhibits a pure $\ell$ = 2 character, the measured distributions indicate strong mixing between $\ell$ = 0 and $\ell$ = 2 transfer for all levels. These observed distributions were then compared with reaction calculations in the Adiabatic Distorted Wave Approximation (ADWA) performed, using the code {\tt TWOFNR} \cite{TWOFNR}. Here, the Johnson-Soper adiabatic model \cite{Johnson} was employed with standard parameters \cite{Lee} using zero range and the Koning-Delaroche \cite{Koning} global nucleon-nucleus optical potential. We estimate an uncertainty in the overall cross section normalization of $\sim$20\%, with the dominant contribution coming overwhelmingly from the modelling of the ($d,p$) reaction itself~\cite{Lee}. The solid angle was calculated accurately from the known geometry, the fitted position of the beam spot and omitting the detector strips excluded from the analysis. The systematic uncertainty in the normalisation of the data arises principally from the uncertainty in the target thickness (taken as 10$\%$) since the total number of incident particles was precisely given by a direct measurement included continuously in the data stream. A summary of the properties of excited states in $^{24}$Ne determined in this work is given in Table \ref{table1}, together with a comparison with our shell-model calculations using the USDA interaction \cite{Brown}. A proposed matching of analog levels in $^{24}$Si is also shown \cite{Wolf}. We have adopted a number of mirror assignments from earlier work \cite{Wolf} and, although the spin-parity assignments of the 3449- and 3471-keV excited states in $^{24}$Si are not uniquely defined, we propose analog matchings to the 2$^{+}_1$, 3871-keV and 4$^{+}_1$, 3962-keV levels in $^{24}$Ne, respectively, based on mirror energy differences. Specifically, a pairing to the 0$^{+}_2$ state would require a very large mirror energy shift of $\sim$1.3 MeV (although we note that a recent study \cite{Longfellow} suggested that such an assignment may be possible).


The present results show excellent agreement with shell model calculations, especially for $\ell$ = 0 transfers. Notably, we find that $\ell$ = 2 strengths for strongly mixed states can deviate considerably from theory, as was previously highlighted in Ref. \cite{Lotay}. This is particularly relevant to the extraction of astrophysical data. Specifically, the authors of Ref. \cite{Wolf} were forced to rely on shell model ratios of $\ell$ = 0 and $\ell$ = 2 strengths in order to analyse their angle-integrated cross sections, but their extracted $\ell$ = 0 values are then susceptible to inaccuracies in the shell model theory (the $\ell$ = 0 strength determines the important resonance parameters for astrophysics). The present work measures the $\ell$ = 2 and $\ell$ = 0 strengths independently of any prior constraints and indeed we find clear differences with the results from Ref. \cite{Wolf}. In particular, the values of $C^2S_{\ell=0}$ of the 2$^+_1$ and $2^+_2$ excited levels in the $T=2$ system are found, respectively, to be 0.28(6) [compared to 0.6(2) for Ref. \cite{Wolf}] and 0.44(9) [compared to 0.7(4)]. The differences in both magnitude and uncertainty have important consequences for the role of the $^{23}$Al($p,\gamma$) reaction in determining the development of X-ray bursters.

%

\begin{figure}[h]
  \includegraphics[width=\linewidth]{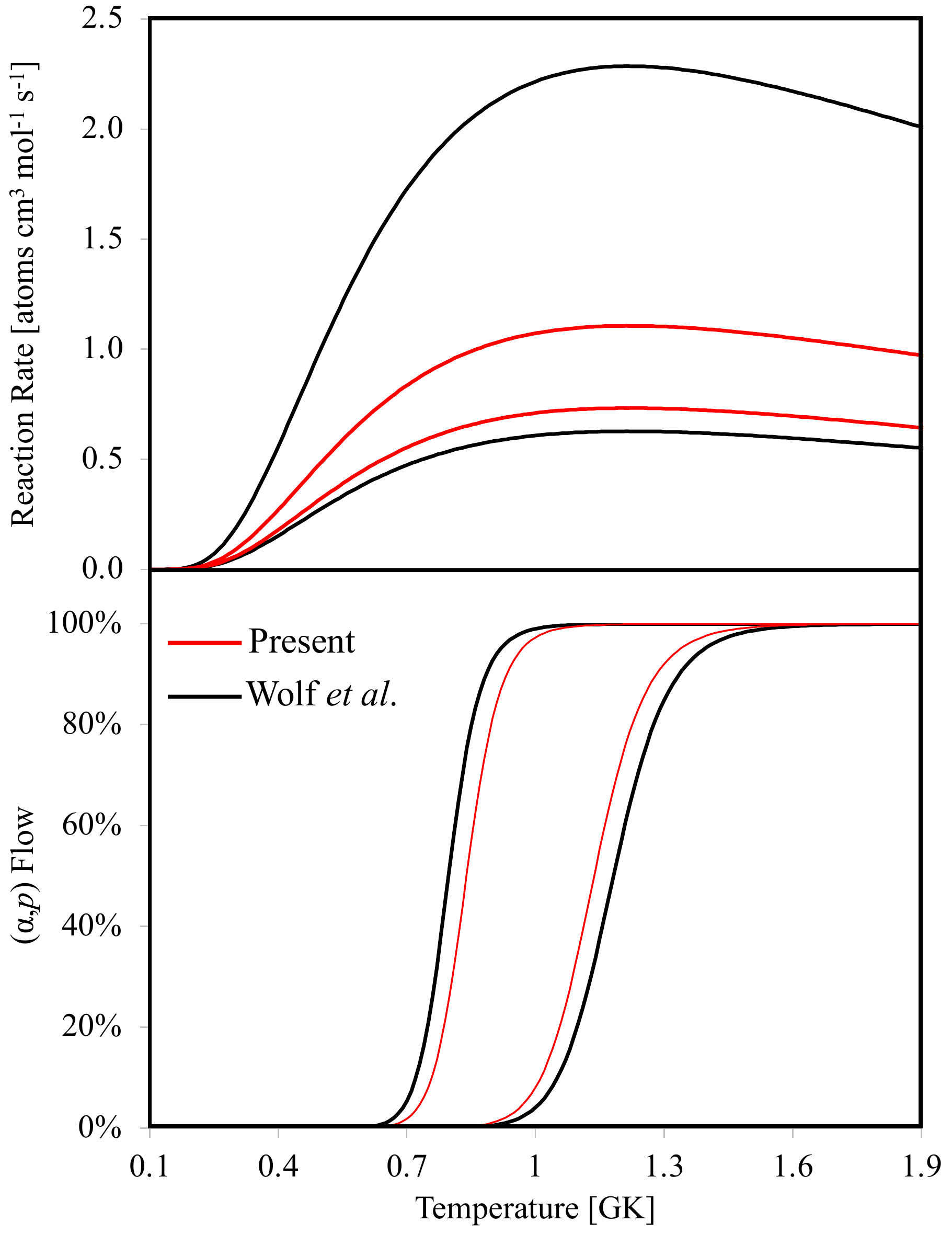}%
\caption{(color online) (top) Contribution of the 157-keV resonance to the $^{23}$Al($p,\gamma$) stellar reaction rate based only on uncertainties in its resonance strength from the present measurement in comparison with those of Ref. \cite{Wolf}. (bottom) Percentage flow of material from the $^{22}$Mg waiting point, through the ($\alpha,p$) process, based on 1.5$\sigma$ uncertainties in the present $^{23}$Al($p,\gamma$) reaction rate in comparison with equivalent 1.5$\sigma$ uncertainties in the previously reported rate of Wolf {\it et al.} \cite{Wolf}. In this case, the contribution of all resonances have been included, as well as uncertainties associated with the reaction Q-value.}
\label{Fig3}
\end{figure}

For an evaluation of the astrophysical $^{23}$Al($p,\gamma$) reaction rate, we consider the contribution of excited states in $^{24}$Si at $E_x$ = 3449, 3471, 4170 and 4470 keV, corresponding to resonances in the $^{23}$Al + $p$ system at $E_r$ = 157, 179, 878 and 1178 keV, respectively (see Table \ref{table2}; the direct capture component is expected to be negligible for temperatures, $T\geq0.1$ GK, and we do not foresee any significant departure from the value previously reported in Ref. \cite{Banu} based on the present results). Here, we adopt spectroscopic factors obtained in the present work for the determination of proton partial widths. The spectroscopic factors of mirror analog states are expected to be nearly identical \cite{Timofeyuk1,Timofeyuk2}. In assessing the validity of this statement, we performed a comparison of proton and neutron spectroscopic factors in the mirror systems: $^{17}$F$-$$^{17}$O \cite{Ajzenberg,Cooper}, $^{21}$Na$-$$^{21}$Ne \cite{Terakawa,Mukhamedzhanov,Howard,Stanford}, $^{25}$Al$-$$^{25}$Mg \cite{Entezami,Berg}, $^{29}$P$-$$^{29}$Si \cite{Dykoski,Mermaz,Peterson}, and $^{33}$Cl$-$$^{33}$S \cite{Mermaz,Morrison,Kozub}, up to excitation energies of $\sim$4 $-$ 5 MeV. We found that spectroscopic factors agree to within $\sim$12$\%$, with a standard deviation of $\sim$10$\%$. This is well within the known $\sim$20$\%$ uncertainty associated with the extraction of spectroscopic factors from experimentally measured cross sections. As such, we conclude that spectroscopic factors obtained for excited states in $^{24}$Ne may be adopted for analog levels in $^{24}$Si to a precision consistent with experimental uncertainties. In contrast, $\gamma$-ray partial widths were calculated using transition densities from our USDA shell-model calculations, adapted to the actual transition energies between the $^{24}$Si states shown in Table \ref{table1}. It should be noted that the values of $\Gamma_{\gamma}$ are negligible for the determination of $\omega$$\gamma$ for the 157- and 179-keV resonances, as they are significantly larger than the corresponding proton partial widths, $\Gamma_p$. However, in the case of the 878- and 1178-keV states, the opposite is true. For the 1178-keV state, the present value of $\Gamma_{\gamma}$ is in good agreement with the USD results of Ref. \cite{Herndl}, whereas our current estimate for the 878-keV resonance is a factor ~2 smaller. In the case of the latter, we note that while there is a discrepancy between the USDA and USD calculations, the contribution of the resonance at 878 keV to the overall $^{23}$Al($p,\gamma$) stellar reaction rate is negligible for temperatures, $T$ = 0.1 $-$ 2 GK.



In agreement with previous studies \cite{Schatz3,Wolf}, we find that the $\ell$ = 0 resonance at 157 keV makes the most significant contribution to the $^{23}$Al($p,\gamma$) stellar reaction rate for $T=0.1-2$ GK. However, in contrast to previous work \cite{Wolf}, uncertainties in the strength of the 157-keV resonance have been reduced by a factor of $\sim$4. Consequently, in order to fully assess the astrophysical implications of the current study, we have estimated the uncertainty in the total reaction rate based on the present resonance energies (which have an uncertainty dominated by the reaction Q-value \cite{Wang}) and the resonance strengths (with uncertainties dominated by the spectroscopic factors, but now much improved). A 1.5$\sigma$ confidence interval was calculated to properly account for experimental uncertainties in the $^{23}$Al($p,\gamma$) reaction parameters, which we note also accounts for uncertainties in the $^{22}$Mg($\alpha,p$) reaction cross section \cite{Randhawa}. We note that the authors of Ref. \cite{Randhawa} utilised the {\tt TALYS} code to extend their data into the Gamow energy window for Type-I X-ray bursts and, as such, we presently estimate a $\sim$60$\%$ uncertainty in the $^{22}$Mg($\alpha,p$) rate $-$ this does not include uncertainties associated with centre-of-mass energies in Ref. \cite{Randhawa}. By using the Saha equation to determine the $^{22}$Mg($p,\gamma$)/$^{23}$Al($p,\gamma$) equilibrium \cite{Schatz4}, and comparing the present results with those of Ref. \cite{Wolf}, we have been able to investigate the relative competition between the $^{22}$Mg($p,\gamma$)$^{23}$Al($p,\gamma$)$^{24}$Si reaction sequence and the $^{22}$Mg($\alpha,p$)$^{25}$Al process path \cite{Randhawa} (assuming ignition conditions of Ref. \cite{Fisker} and total accreted mass fractions consistent with the ``zM'' model of Ref. \cite{Woosley}). In particular, in defining the temperature at which the $^{22}$Mg($\alpha,p$) reaction governs 50$\%$ of the nucleosynthetic flow in Type-I X-ray bursts as the ``tipping" point between the $rp$- and ($\alpha,p$) processes, we find that the latter will only become significant at temperatures $\gtrsim$0.85 GK, as shown in Fig. \ref{Fig3}. Such temperatures are only briefly reached for standard X-ray burst model calculations \cite{Fisker,Merz} and, in ruling out the previously possible lower-temperature onset of the ($\alpha,p$) process \cite{Randhawa,Wolf}, we may now conclude that the pathway through the $^{22}$Mg($\alpha,p$) reaction is not relevant for anything but the most energetic bursters.

In summary, we have performed the first measurement of the $^{23}$Ne($d,p$)$^{24}$Ne transfer reaction. Several strong single-particle states in $^{24}$Ne have been identified and their associated neutron spectroscopic factors extracted to a precision of $\sim$20$\%$. Using these spectroscopic factors to deduce the properties of resonant states in the astrophysical $^{23}$Al($p,\gamma$) reaction, we have reduced uncertainties in the strength of the key $E_r$ = 157 keV, $\ell$ = 0 level, in comparison with the most recent study of Ref. \cite{Wolf}, by a factor of $\sim$4, considerably constraining the rate over the temperature range of X-ray bursts. In particular, we find that the $^{23}$Al($p,\gamma$)$^{24}$Si reaction is effective in bypassing the $^{22}$Mg waiting point in the $rp$ process (according to standard modelling conditions) for temperatures up to at least 0.85 GK, while the $^{22}$Mg($\alpha,p$) pathway might play a more prevalent role above 1 GK, the very peak temperature region only rarely reached in X-ray bursts. Further constraints on the $^{23}$Al($p,\gamma$) reaction would now require a precise determination of the reaction Q-value \cite{Wang} and, in this regard, we understand that a new measurement of the $^{24}$Si mass was recently performed at the National Superconducting Cyclotron Laboratory, USA \cite{Puentes}. The results for resonance strengths, combined with a precise Q-value determination, are now likely to constrain the uncertainties in the nuclear physics data sufficiently tightly to allow the accurate extraction of neutron star mass-radius ratios from current experimental observations of Type-I X-ray bursts \cite{Meisel}.


{\it Acknowledgements} -- The authors would like to thank the TRIUMF beam delivery group for their efforts in providing high-quality beams. This work has been supported by the Natural Sciences and Engineering Research Council of Canada (NSERC), The Canada Foundation for Innovation and the British Columbia Knowledge Development Fund. TRIUMF receives federal funding via a contribution agreement through the National Research Council of Canada. UK personnel were supported by the Science and Technologies Facilities Council (STFC), JH acknowledges support at the University of Surrey under UKRI Future Leaders Fellowship MR/T022264/1,  MR acknowledges support from the the U. S. Department of Energy, Office of Science through grant No. DE-SC0016988 and ZM acknowledges support from the U.S. Department of Energy under grants DE-FG02-88ER40387 and DESC001904. We also benefited from support by the National Science Foundation under grant PHY-1430152 (Joint Institute for Nuclear Astrophysics$-$Center for the Evolution of the Elements).

 \normalsize


\begin{thebibliography}{99}
 
\bibitem{Wallace} R. K. Wallace and S. E. Woosley, Astrophys. J. {\bf45}, 389 (1981).
\bibitem{Schatz} H. Schatz and K.E. Rehm, Nucl. Phys. A{\bf777}, 601 (2006).
\bibitem{Schatz2} H. Schatz {\it et al.}, Phys. Rev. Lett. {\bf 86}, 3471 (2001).


\bibitem{Caughlan} G. R. Caughlan and W. A. Fowler, Nature Phys. Sci. {\bf 238}, 23 (1972).
\bibitem{Audouze} J. Audouze, J. W. Truran and B. A. Zimmerman, Astrophys. J. {\bf 184}, 493 (1974).
\bibitem{Fisker} J. L. Fisker, H. Schatz and F-K. Thielemann, Astrophys. J. Suppl. Ser. {\bf 174}, 261 (2008).


\bibitem{Woosley} S. E. Woosley {\it et al.}, Astrophys. J. Suppl. Ser. {\bf 151}, 75 (2004).
\bibitem{Paxton} B. Paxton {\it et al.}, Astrophys. J. Suppl. Ser. {\bf 220}, 15 (2015).
\bibitem{Jose} J. Jos{\'e}, F. Moreno, A. Parikh and C. Iliadis, Astrophys. J. Suppl. Ser. {\bf 189}, 204 (2010).




\bibitem{Parikh} A. Parikh, J. Jos{\'e}, F. Moreno and C. Iliadis, Astrophys. J. Suppl. Ser. {\bf 178}, 110 (2008).
\bibitem{Cyburt} R. H. Cyburt {\it et al.}, Astrophys. J. {\bf 830}, 2 (2016).

\bibitem{Meisel} Z. Meisel, G. Merz and S. Medvid, Astrophys. J. {\bf 872}, 84 (2019).



\bibitem{Meisel2} Z. Meisel, Astrophys. J. {\bf 860}, 147 (2018).

\bibitem{Randhawa} J. S. Randhawa {\it et al.}, Phys. Rev. Lett. {\bf 125}, 202701 (2020).

\bibitem{Galloway} D. K. Galloway, M. P. Muno, J. M. Hartman, D. Psaltis and D. Chakrabarty, Astrophys. J. {\bf 179}, 360 (2008). 
\bibitem{Galloway2} D. K. Galloway {\it et al.}, Astrophys. J. Suppl. Ser. {\bf 249}, 32 (2020).



\bibitem{Schatz3} H. Schatz {\it et al.}, Phys. Rev. Lett. {\bf 79}, 3845 (1997).
\bibitem{Yoneda} K. Yoneda {\it et al.}, Phys. Rev. C {\bf 74}, 021303(R) (2006).
\bibitem{Banu} A. Banu {\it et al.}, {\it et al.}, Phys. Rev. C {\bf 86}, 015806 (2012). 

\bibitem{Wolf} C. Wolf {\it et al.},  Phys. Rev. Lett. {\bf 122}, 232701 (2019).
\bibitem{Wang} M. Wang {\it et al.}, Chin. Phys. C {\bf 41}, 3 (2017).

\bibitem{Lotay} G. Lotay {\it et al.}, Eur. Phys. J. A {\bf 56}, 3 (2020).

\bibitem{Iliadis} C. Iliadis, P. M. Endt, N. Prantzos, and W. J. Thompson, Astrophys. J. {\bf 524}, 434 (1999).
\bibitem{Pain} S. D. Pain {\it et al.}, Phys. Rev. Lett. {\bf 114}, 212501 (2015).
\bibitem{Margerin} V. Margerin {\it et al.}, Phys. Rev. Lett. {\bf 115}, 062701 (2015).

\bibitem{TIGRESS} G. Hackman and C. E. Svensson, Hyperfine Interact. {\bf 225}, 241 (2014).
\bibitem{SHARC} C. Aa. Diget {\it et al.}, J. Instrum. {\bf 6} PO2005 (2011).

\bibitem{Wilson} G. L. Wilson {\it et al.}, Phys. Lett. B {\bf 759}, 417 (2016).
\bibitem{Matta} A. Matta {\it et al.}, Phys. Rev. C {\bf 99}, 044320 (2019).

\bibitem{Bragg} S. Cruz {\it et al.}, Phys. Lett. B {\bf 786}, 94 (2018).


\bibitem{Brown} B. A. Brown and W. A. Richter, Phys. Rev. C {\bf 74}, 034315 (2006).



\bibitem{Hoffman} C.R. Hoffman {\it et al.}, Phys. Rev. C {\bf 68}, 034304 (2003).

\bibitem{Herndl} H. Herndl, J. G{\"o}rres, M. Wiescher, B. A. Brown and L. Van Wormer, Phys. Rev. C {\bf 52}, 2 (1995).

\bibitem{Brown2} B. A. Brown and W. D. M. Rae, Nucl. Dat. Sheets {\bf 120}, 115 (2014).


\bibitem{TWOFNR} J. A. Tostevin, University of Surrey version of the code TWOFNR (of M. Toyama, M. Igarashi and N. Kishida) and code FRONT (private communication), http://www.nucleartheory.net/NPG/code.htm.

\bibitem{Johnson} R. C. Johnson and P. J. R. Soper, Phys. Rev. C {\bf 1}, 976 (1970).

\bibitem{Lee} J. Lee {\it et al.}, Phys. Rev. C {\bf 75} 064320 (2007).


\bibitem{Koning} A. J. Koning and J. P. Delaroche, Nucl. Phys. A {\bf 713}, 231 (2003).


\bibitem{Longfellow} B. Longfellow {\it et al.}, Phys. Rev. C {\bf 101}, 031303(R) (2020).








\bibitem{Timofeyuk1} N. K. Timofeyuk, R. C. Johnson, and A. M. Mukhamedzhanov, Phys. Rev. Lett. {\bf 91}, 232501 (2003).
\bibitem{Timofeyuk2} N. K. Timofeyuk, P. Descouvemont, and R. C. Johnson, Eur. Phys. J. A {\bf 27}, 269 (2006).




\bibitem{Ajzenberg} F. Ajzenberg-Selove, Nucl. Phys. A {\bf 375}, 1 (1982).
\bibitem{Cooper} M. D. Cooper, W. F. Hornyak and P. G. Roos, Nucl. Phys. A {\bf 218}, 249 (1974).

\bibitem{Terakawa} A. Terakawa {\it et al.}, Phys. Rev. C {\bf 48}, 2775 (1993).
\bibitem{Mukhamedzhanov} A. M. Mukhamedzhanov {\it et al.}, Phys. Rev. C {\bf 73}, 035806 (2006).
\bibitem{Howard} A. J. Howard, J. O. Pronko and C. A. Whitten Jr., Nucl. Phys. A {\bf 152}, 317 (1970).

\bibitem{Stanford} A.R. Stanford and P. A. Quin, Nucl. Phys. A {\bf 342}, 283 (1980).

\bibitem{Entezami} F. Entezami, A. K. Basak, O. Karban, P. M. Lewis and S. Roman, Nucl. Phys. A {\bf 366}, 1 (1981).
\bibitem{Berg} G. P. A. Berg, R, Das, S. K. Datta and P. A. Quin, Nucl. Phys. A {\bf 289}, 15 (1977).

\bibitem{Dykoski} W. W. Dykoski and D. Dehnhard, Phys. Rev. C {\bf 13}, 80 (1976).
\bibitem{Peterson} R. J. Peterson, C. A. Fields, R. S. Raymond, J. R. Thieke and J. L. Ullman, Nucl. Phys. A {\bf 408}, 221 (1983).
\bibitem{Mermaz} M. C. Mermaz {\it et al.}, Phys. Rev. C {\bf 4}, 1778 (1971).


\bibitem{Morrison} R. A. Morrison, Nucl. Phys. A {\bf 140}, 97 (1970).
\bibitem{Kozub} R. L. Kozub, Phys. Rev. C {\bf 5}, 413 (1972).



\bibitem{Schatz4} H. Schatz {\it et al.}, Phys. Rep. {\bf 294}, 167 (1998).



\bibitem{Merz} G. Merz and Z. Meisel, Mon. Not. Astron. Soc. {\bf 500}, 2958 (2021).


\bibitem{Puentes} D. Puentes {\it et al.}, {\it private communication}.


 \end{thebibliography}
 \end{document}